\documentclass[doublecol]{epl2}
\usepackage{graphicx,epsfig,color}

\title{Convective Ripening and Initiation of Rainfall}

\author{Michael Wilkinson}

\institute{
{Department of Mathematics and Statistics,
The Open University, Walton Hall, Milton Keynes, MK7 6AA, England}
}
\pacs{92.60.Nv}{Cloud physics and chemistry}
\pacs{92.60.hk}{Convection, turbulence, and diffusion}
\pacs{92.60.Mt}{Particles and aerosols}

\abstract{
This paper discusses the evolution of the droplet size distribution
for a liquid-in-gas aerosol contained in a Rayleigh-B\'enard cell.
It introduces a non-collisional model for broadening the droplet size
distribution, termed \lq convective ripening'. The paper also considers
the initiation of rainfall from ice-free cumulus clouds. It is argued
that while collisional mechanisms cannot explain the production of rain from
clouds with water droplet diameters of $20\ \mu {\rm m}$, the non-collisional
convective ripening mechanism gives a much faster route to increasing the size
of the small fraction of droplets that grow into raindrops.
}

\begin{document}

\maketitle

\section{Introduction}
\label{sec: 1}

The dynamics of the onset of rainfall
from ice-free (\lq warm') cumulus clouds is poorly understood \cite{Mas57,Pru+97,Rog+82,Sha03}.
Coalescence of droplets which collide due to differential rates of gravitational
settling is effective for droplets with radius $a$ above $50\mu{\rm m}$, and leads
to a runaway growth to produce millimetre-scale raindrops \cite{Kos+05}.
Many clouds are found to contain droplets with radius approximately
$10-15\mu{\rm m}$, which result from primary condensation onto
aerosol nuclei. For droplets in this size range, growth
by collisional coalescence is slow because the collision rates
and the collision efficiencies are low \cite{Mas57}. This makes it difficult to
explain observations of the rapid onset of rainfall from warm cumulus clouds. (Rainfall
from ice-bearing clouds is easier to explain: see \cite{Mas57} for a discussion
of the Bergeron process).

It is, therefore, desirable to formulate models for non-collisional growth
of water droplets, in which some droplets are able to grow at the expense
of others shrinking, by transferring water molecules between droplets as
water vapour. Ostwald ripening \cite{Lif+61} is one such mechanism,
but it is too slow to be significant in terrestrial clouds \cite{Cle08},
while it is relevant to test-tube models for rainfall \cite{Wil14}.
It has been suggested that condensation processes may
be able to cause the droplet size distribution to broaden
due to fluctuations in the degree of supersaturation. This
possibility has been addressed by numerous authors: see, for example,
\cite{Bre+01,Cha+01,Vai+01,Vai+02,And+04,Cel+05,Sid+09,Lan+09,Kum+13}.
These investigations have used numerical simulations, and it is
difficult to draw conclusions which are applicable to real clouds
because of the limited range of size scales which can be simulated reliably.
The models which are used in these studies also have a large
number of parameters. These factors make it difficult to obtain
general conclusions.

This work will consider a benchmark model for the
broadening of the droplet size distribution
of an aerosol due to convection. This process will be termed
\lq convective ripening' to distinguish it from Ostwald ripening.
This work considers how the process works
in the simplest relevant model, which is an aerosol in a
Rayleigh-B\'enard convection cell. As well as having fewer physical
parameters than a cloud, this system can be subject to a carefully controlled
laboratory investigation. It would, however, be very difficult to perform a
numerical simulation of this system in the parameter range describing
clouds, because the important physical processes involve
all lengthscales of the system.

Having described a non-collisional model for droplet growth,
this will be applied to rain initiation from ice-free
(\lq warm') cumulus clouds. An important aspect of this problem is that
the conversion of a microscopic water droplet into a rain droplet is a
very rare event (this fact was previously emphasised by Kostinski and
Shaw \cite{Kos+05}). However, the growth of a droplet to the stage where
runaway growth occurs is a multi-stage process. It is argued that the probability
for the required number of favourable events to be achieved by collisional
processes alone is extremely small. It is shown that droplets can grow
much more rapidly the convective ripening mechanism.

This paper complements a recent work which discussed Ostwald ripening
as a non-collisional model for the evolution of the droplet size distribution
\cite{Wil14}. That work concluded that while Ostwald ripening provides
the correct description of a test-tube model for rainfall, it acts too slowly
to explain rain from warm cumulus clouds.

Discussions of convection processes within clouds often
involve the complex and poorly understood
issue of \lq entrainment' of air into a cloud (see, for example
\cite{And+04,Kum+13}). This paper argues that a Rayleigh-B\'enard
cell appears to be a sufficient model to understand the mechanism of rainfall
from warm cumulus clouds. Because entrainment does
not enter into the Rayleigh-B\'enard system which is
discussed in this paper, hypotheses about entrainment are not a necessary feature
of understanding rainfall from warm cumulus clouds.

\section{The convection cell model}
\label{sec: 2}

Consider a Rayleigh-B\'enard convection cell, in which the working
fluid is a gas (air, say) containing an aerosol suspension of liquid droplets
(water, say). The height of the cell is $h$ and the temperature
difference between the upper and lower plates is $\Delta T_h$. It is
assumed that the horizontal dimensions of the cell are large compared
to $h$. The rate of heat transfer per unit area is $Q$. The gas has volume-specific
heat capacity at constant pressure $C_{\rm g}$,
density $\rho_{\rm g}$, kinematic viscosity $\nu$ and thermal
diffusivity is $D_{\rm th}$.
It will be assumed the convection in the container is in a turbulent regime,
with rate of dissipation per unit mass $\epsilon$.

The cell contains an aerosol of liquid droplets with density $\rho_{\rm l}$ and
volume-specific heat of evaporation $L$. The vapour of the
aerosol liquid in the carrier gas has diffusivity $D$.
The number density of droplets is $n_0$, and the probability density function for the
droplet radius $a$ at time $t$ is $P(a,t)$.

The objective is to understand how convection affects the distribution
of sizes of the aerosol droplets. It will be assumed that the rate of collisions
between the droplets is negligible. It will also be assumed that collisions of
aerosol droplets with the walls of the container is not a significant process.
The validity of this assumption is not critical to using this system as a model
for cloud physics (because there is no material container in that context).

In order to understand the ripening of the droplet size distribution it is
necessary to consider first how the aerosols responds to changes in the temperature
of the surrounding gas, and then how the convection process influences the temperature.

\section{Response to temperature fluctuations}
\label{sec: 3}

Changes in the temperature of the surrounding gas cause the
size of the droplets to change due to condensation or evaporation.
This is characterised by two parameters, $T_0$ and $\tau_{\rm eq}$,
which describe, respectively, the sensitivity and the timescale of the
response. It will be shown that if the majority of the aerosol droplets
have radius close to $a_0$, the change $\delta a_0$ of
the equilibrium radius in response
to a temperature increment $\delta T$ satisfies
\begin{equation}
\label{eq: 3.1}
\frac{\delta a_0}{a_0}=-\frac{\delta T}{T_0}
\end{equation}
to leading order in $\delta T$, and that the change in droplet radius occurs on a
timescale $\tau_{\rm eq}$. In the following expressions for both $T_0$
and $\tau_{\rm eq}$ are obtained. Equivalent calculations can be found in many earlier works (reviewed in
\cite{Mas57,Pru+97,Rog+82,Sha03}), but with differences in physical motivation an notation. A brief
derivation is given here to make this paper unambiguous and self-contained.

The volume fraction of water molecules in the gas, $\Phi$, may be assumed to be uniform
throughout the container because the system is well mixed by convection.
This is the sum of contributions from water in the liquid and the vapour
phase:
\begin{equation}
\label{eq: 3.2}
\Phi=\Phi_{\rm l}+\Phi_{\rm v}
\ .
\end{equation}
The equilibrium vapour content above a flat liquid surface at temperature $T$
is denoted by $\Phi_{\rm eq}(T)$, and there may be a degree of supersaturation,
denoted by $s$. It is assumed that the droplets are sufficiently large that
curvature and hygroscopic effects of the aerosol
condensation nuclei can be neglected, so that the vapour mole fraction
in the bulk of the gas phase will be written
\begin{equation}
\label{eq: 3.3}
\Phi_{\rm v}=\Phi_{\rm eq}(T)+s
\ .
\end{equation}
The volume-fraction of the liquid phase is
\begin{equation}
\label{eq: 3.4}
\Phi_{\rm l}=\frac{4\pi}{3}n_0\langle a^3\rangle
\end{equation}
(throughout this paper $\langle X\rangle$ denotes the expectation value
of any quantity $X$).
If the temperature of the gas changes, the sizes of the droplets will change.
For example, a decrease of the temperature results in a supersaturation which
causes condensation on the surface of the droplets. The rate of the condensation
process is determined by diffusion of vapour. The radius of a droplet changes at a rate
\begin{equation}
\label{eq: 3.5}
\frac{{\rm d}a}{{\rm d}t}=j_{\rm v}=-\frac{D}{a}\Delta \Phi_{\rm v}
\end{equation}
where $j_{\rm v}$ is the volume flux density of condensing
molecules and $\Delta \Phi_{\rm v}$ is the the volume-fraction
on the surface of the droplet minus the volume-fraction in the bulk
of the gas phase. The surface of the droplet is in quasi-static
equilibrium with the surrounding fluid, so there is no supersaturation
at the surface. However, the temperature of the liquid droplet may be
increased by an amount $\Delta T$ due to the latent heat of water
condensing on the surface, so that
\begin{equation}
\label{eq: 3.6}
\Delta \Phi_{\rm v}=\frac{{\rm d}\Phi_{\rm eq}}{{\rm d}T}\Delta T-s
\ .
\end{equation}
The thermal flux density due to the latent heat is
\begin{equation}
\label{eq: 3.7}
j_{\rm th}=-L\frac{{\rm d}a}{{\rm d}t}=-\frac{D_{\rm th}C_{\rm g}}{a}\Delta T
\ .
\end{equation}
Combining (\ref{eq: 3.6}) and (\ref{eq: 3.7}) gives
\begin{equation}
\label{eq: 3.8}
\frac{{\rm d}a}{{\rm d}t}=\frac{D}{a}s-
\frac{D}{a}\frac{{\rm d}\Phi_{\rm eq}}{{\rm d}T}\frac{La}{D_{\rm th}C_{\rm g}}
\frac{{\rm d}a}{{\rm d}T}
\ .
\end{equation}
This gives a simple expression relating the rate of droplet
growth to the supersaturation
\begin{equation}
\label{eq: 3.9}
\frac{{\rm d}a}{{\rm d}t}=\frac{D_{\rm eff}}{a}s
\end{equation}
where the effective diffusion constant is
\begin{equation}
\label{eq: 3.10}
D_{\rm eff}=\frac{D}{1+\Theta}
\ ,\ \ \
\Theta=\frac{DL}{D_{\rm th}C_{\rm g}}\frac{{\rm d}\Phi_{\rm eq}}{{\rm d}T}
\ .
\end{equation}
This treatment neglected the possibility of cross-coupling
between thermal and mass fluxes (the Soret effect). This could be incorporated with a simple
modification of the theory, but the coefficients of the off-diagonal terms
of the transport matrix do not appear to have been definitively determined.

In the case where the temperature of the system varies extremely slowly, the
supersaturation is always negligible, and the relation between droplet size
and temperature is determined by writing
\begin{equation}
\label{eq: 3.11}
\Phi=\frac{4\pi}{3}n_0a_0^3(T)+\Phi_{\rm eq}(T)
\end{equation}
so that a small change in temperature $\delta T$ results in a small change of radius
$\delta a_0$ given by (\ref{eq: 3.1}), with coefficient
\begin{equation}
\label{eq: 3.12}
T_0=3 \Phi_{\rm l} \left(\frac{{\rm d}\Phi_{\rm eq}}{{\rm d}T}\right)^{-1}
\ .
\end{equation}
Now consider the effect of varying the temperature of the aerosol at a
finite rate, but still assuming that it is spatially homogeneous: write
$T=T_0+\delta T(t)$, $a=a_0(T_0)+\delta a(t)$, so that
\begin{equation}
\label{eq: 3.13}
\Phi=\frac{4\pi}{3}n_0(a_0+\delta a)^3+\Phi_{\rm eq}(T_0)
+\frac{{\rm d}\Phi_{\rm eq}}{{\rm d}T}\delta T(t)+s
\ .
\end{equation}
Taking the leading order in the small fluctuation $\delta a$ and using
(\ref{eq: 3.9}) yields the following equation for the response of the
droplets to fluctuations in temperature:
\begin{equation}
\label{eq: 3.14}
\frac{{\rm d}\delta a}{{\rm d}t}=-\frac{1}{\tau_{\rm eq}}\delta a
-\frac{a_0}{\tau_{\rm eq}T_0}\delta T(t)
\end{equation}
where the relaxation time is
\begin{equation}
\label{eq: 3.15}
\tau_{\rm eq}=\frac{a_0^2}{3\Phi_{\rm l}D_{\rm eff}}=\frac{1}{4\pi n_0a_0D_{\rm eff}}
\ .
\end{equation}
At this stage it is relevant to make some estimates
of the parameters $T_0$ and $\tau_{\rm eq}$. The rate of change of the
saturation volume-fraction is obtained from the Clausius-Clapeyron
relation, ${\rm d}p/{\rm d}T=L/T\Delta V$, where $\Delta V$ is the volume
change on a phase transition. Assuming that the vapour pressure
is sufficiently low that the ideal gas law is applicable,
\begin{equation}
\label{eq: 3.16}
p=\frac{RT}{V_{\rm m}}\Phi_{\rm eq}
\ ,
\end{equation}
where $V_{\rm m}$ is the molar volume of the
liquid. Also, the volume change per mole associated with the phase
transition is $\Delta V=V_{\rm m}/\Phi_{\rm eq}$.
The Clausius-Clapeyron equation can therefore be written
in the form
\begin{equation}
\label{eq: 3.17}
\frac{{\rm d}p}{{\rm d}T}=\frac{L}{T}\Phi_{\rm eq}
\ .
\end{equation}
Comparing (\ref{eq: 3.16}) and (\ref{eq: 3.17}) yields an
expression for ${\rm d}\Phi_{\rm eq}/{\rm d}T$, and hence
\begin{equation}
\label{eq: 3.18}
T_0=\frac{3RT}{LV_{\rm m}-RT}\frac{\Phi_{\rm l}}{\Phi_{\rm eq}}T
\ .
\end{equation}
For water at $T=278\,{\rm K}$, $L=2.4\times 10^9\,{\rm J}\,{\rm m}^{-3}$, $C_{\rm g}=800\,{\rm Jm}^{-3}$,
$V_{\rm m}=1.8\times 10^{-5}\,{\rm m}^3{\rm mol}^{-1}$, $\Phi_{\rm eq}=7\times 10^{-6}$, $D=2.5\times 10^{-5}\,{\rm m^2}{\rm s}^{-1}$
and $D_{\rm th}=1.9\times 10^{-5}\,{\rm m}^2{\rm s}^{-1}$. These data yield $\Theta\approx 1$. If the liquid water content
is $10\%$ of the total water content, then $T_0\approx 5\,{\rm K}$: that is, the droplet
size is very sensitive to changes of temperature.
If the droplets are of size $a=10\rm \mu{\rm m}$ and density $n_0=4\times 10^8\,{\rm m}^{-3}$,
(which are typical values for clouds) the equilibration time is $\tau_{\rm eq}=1.6\,{\rm s}$.

\section{Ripening in a turbulent convection cell}
\label{sec: 4}

Now consider the response of the aerosol to convective motion
in the cell. This is a consequence of how the temperature changes along
the trajectories of the aerosol droplets (which are assumed to be advected by the flow).

Turbulent convection in a Rayleigh-B\'enard cell is reviewed in \cite{Sig94,Bod+00,Ahl+09}.
The upper and lower plates are at temperatures $T_{\rm up}$ and $T_{\rm low}$
respectively. The expectation value of the temperature is close to
$T_{\rm av}=(T_{\rm up}+T_{\rm low})/2$ (with
small logarithmic corrections) \cite{Ahl+12}, except in the vicinity of the upper
and lower plates, and at any time most of the gas in the convection cell
is at a temperature close to $T_{\rm av}$.
Gas which is in contact with the lower plate of the cell is heated to a
higher temperature $T_{\rm av}+\Delta T$
(where $\Delta T\le \Delta T_h/2$), and joins a plume of rising gas.
The plumes persist on a timescale $\tau_{\rm c}$, which cannot exceed that of the largest eddies,
$\tau_h=(h^2/\epsilon)^{1/3}$. It will be assumed that $\tau_{\rm c}/\tau_{\rm eq}\gg 1$.
However, the mixing process is highly discontinuous.
The plumes form fronts and later tendrils of approximately homogenous gas,
which remain at a temperature close to the temperature that they had upon separation
from the top or bottom plate until the last stage of the mixing process.
In the final stage of mixing a tendril formed by the plume mixes rapidly
with gas from the interior of the cell, which is at a temperature close to $T_{\rm av}$.
The final stage of mixing occurs on a much
shorter timescale, namely the Kolmogorov timescale,
$\tau_{\rm K}=\sqrt{\nu/\epsilon}$. The equilibration timescale will be
assumed to lie between the timescales describing the
flow: $\tau_h\gg \tau_{\rm eq}\gg \tau_{\rm K}$.

The consequence of this picture is that droplets in the a rising plume are at a
temperature, $T_{\rm av}+\Delta T$, and while the plume forms they equilibrate
to a smaller radius,
\begin{equation}
\label{eq: 4.1}
a_{1-}=a_0-\Lambda \Delta T
\ ,\ \ \
\Lambda \equiv \frac{a_0}{T_0}
\ .
\end{equation}
The gas in the plume rises, without
cooling due to heat exchange, until it reaches the interior of the cell. After a timescale
$\tau_{\rm c}$, the gas in the plume starts to mix with the gas in the interior.
This mixing happens on a timescale which is short compared to the phase
equilibration time, so that droplets of size $a_{1-}$ are mixed
with the droplets in the bulk, which are of size $a_0$. Similarly, plumes of cold
gas which form on the upper plate at a temperature $T_{\rm av}-\Delta T$
inject larger droplets, of radius $a_{1+}=a_0+\Lambda \Delta T$,
when they fall into the bulk. The final stage of this
equilibration happens on a timescale of the Kolmogorov time, which is small
compared to the time required for aerosol droplets to come into equilibrium.
It follows that while the temperature fluctuations associated with the
plume are dissipated, fluctuations in the droplet size remain \lq frozen in',
resulting in a broadening of the droplet size distribution.

Now consider how this model is used to model the evolution of the droplet size
distribution, $P(a,t)$. The plumes carrying gas away
from the lower plate have a distribution of temperature $\Delta T$.
Let $J(\Delta T)\,{\rm d}\Delta T$ be the volume
of gas per unit area, per unit time, which rises from
the lower plate and which has a temperature change
in the interval $[\Delta T,\Delta T+{\rm d}\Delta T]$. It is assumed that
the flux from the upper plate may be described by the same function $J(\Delta T)$.
The material in this temperature range occupies a volume
fraction of the gas in the column equal to
\begin{equation}
\label{eq: 4.2}
\frac{{\rm d}V}{V}=\frac{J(\Delta T)}{h}{\rm d}\Delta T
\ .
\end{equation}
The droplets in this volume fraction undergo a change of radius
equal to $\Delta a=-\Lambda \Delta T$. This results in a change of the
droplet size distribution which satisfies
\begin{equation}
\label{eq: 4.3}
\frac{\partial P}{\partial t}(a,t)=\int_{-\infty}^\infty {\rm d}a'\ {\cal K}(a,a')\,P(a',t)
\end{equation}
where the kernel may be approximated by ${\cal K}(a,a')=K(a-a')$ with
\begin{equation}
\label{eq: 4.4}
K(\Delta a)=\frac{1}{\Lambda h}[J(\Delta a/\Lambda)+J(-\Delta a/\Lambda)]
-\frac{1}{\tau_{\rm c}}\delta (\Delta a)
\ .
\end{equation}
Here $\tau_{\rm c}$ is an estimate for the timescale of a convection roll:
\begin{equation}
\label{eq: 4.5}
\frac{1}{\tau_{\rm c}}=\frac{2}{h}\int_0^\infty {\rm d}x \ J(x)
\ .
\end{equation}
The time $\tau_{\rm c}$ cannot exceed the integral timescale of the flow:
$\tau_{\rm c}\le \tau_{\rm h}=(h^2/\epsilon)^{1/3}$.
The initial rate of broadening of the particle size distribution can be related to the
heat flux in the cell, which is
\begin{equation}
\label{eq: 4.6}
Q=C_p\int_0^\infty {\rm d}\Delta T\ \Delta T\ J(\Delta T)
\end{equation}
where $C_p$ is the specific heat capacity.
The growth of the mean of the absolute value of the size change is
\begin{equation}
\label{eq: 4.7}
\langle |\Delta a|\rangle=\int_{-\infty}^\infty {\rm d}\Delta a\ |\Delta a|\ P(a_0+\Delta a,t)
=\frac{2\Lambda}{hC_p}\,Q\,t
\ .
\end{equation}

It has been argued that the model predicts that droplets change size
discontinuously, in steps with a magnitude comparable
to $\delta a_{\rm max}=\Lambda \Delta T_h/2$.
The typical timescale separating
these jumps of the particle size is
$\tau_{\rm c}\le \tau_h$, but the separation of the steps is random,
and some droplets may experience
several jump events in quick succession.

The production of rain from clouds depends upon droplets reaching a
size which is significantly larger than their original size. In the context of
the Rayleigh-B\'enard model, this would require
a droplet to undergo repeated encounters with the cold plate.
Equation (\ref{eq: 3.9}) implies that the change in surface area of a
droplet is independent of its size. This implies that the
growth in the particle radius after $N$ successive encounters with the upper
plate satisfies
\begin{equation}
\label{eq: 4.8}
\langle a_N^2\rangle =a_0^2+N\Delta A
\end{equation}
for some constant $\Delta A$.

\section{The problem of warm rain initiation}
\label{sec: 5}

A cloud contains water droplets formed by condensation onto microscopic nuclei such as
salt granules, dust grains, or particles of organic matter.
Their concentration and droplet radius are quite variable, but the remainder
of this paper uses the following representative values for a convecting cumulus
cloud which could produce precipitation. The typical droplet radius is $a_0=10\,\mu {\rm m}$,
the number density is $n_0=4\times 10^8\,{\rm m}^{-3}$, and the cloud depth is $h=10^3\, {\rm m}$.
The rate of decrease of temperature with height (lapse rate) is
$3^\circ {\rm C}$ per $1000\,{\rm ft}$, which exceeds the adiabatic lapse rate by $1^\circ {\rm C}$
per $1000\,{\rm ft}$, so that the effective temperature difference between the top and bottom
of the cloud is $3\,{\rm K}$. The typical vertical velocity of air inside the cloud has magnitude
$2\,{\rm m}\,{\rm s}^{-1}$, so that the eddy turnover time may be taken to be $\tau_h=10^3\,{\rm s}$.
An estimate for the rate of dissipation is $\epsilon\approx h^2/\tau_h^3=10^{-3}\,{\rm m}^2{\rm s}^{-3}$,
which gives an estimate of the Kolmogorov time $\tau_{\rm K}\approx 10^{-1}\,{\rm s}$.
Rain falls as droplets of size approximately
$a=1\,{\rm mm}$. A rate of rainfall of
$3.6\,{\rm mm}\,{\rm hr}^{-1}=10^{-6}\,{\rm m}\,{\rm s}^{-1}$ is
described as \lq moderate to heavy rainfall'.

The collision efficiencies $\varepsilon$ of small droplets are somewhat uncertain, but it
is widely accepted that they are low \cite{Mas57,Pru+97}. If the larger droplet has radius
below $20\,\mu{\rm m}$, it is believed that $\varepsilon \le 0.1$, and that for radius $10\,\mu {\rm m}$,
$\varepsilon \le 0.03$ \cite{Pru+97}. For droplets of size
$a=50\,\mu {\rm m}$ colliding with droplets of size $a=10\,\mu {\rm m}$, however,
the efficiencies are expected to be close to unity \cite{Pru+97,Mas57}.

Collisions between droplets settling at a different rate yield a very small
collision rate. The Stokes law for the drag on a sphere at low Reynolds number
indicates that the gravitational settling rate is
\begin{equation}
\label{eq: 5.1}
v=\kappa a^2
\ ,\ \ \
\kappa=\frac{2}{9}\frac{\rho_{\rm l}}{\rho_{\rm g}}\frac{g}{\nu}
\ .
\end{equation}
Inserting values for air and water at $5^\circ {\rm C}$ gives
$\kappa\approx 1.4\times 10^8\,{\rm m}^{-1}{\rm s}^{-1}$.
The collision rate of a droplet of radius $a+\Delta a$ with a gas
of particles of radius $a$ is
\begin{equation}
\label{eq: 5.2}
{\cal R}=4\pi \varepsilon n_0 a^2 \kappa [(a+\Delta a)^2-a^2]\sim 8\pi \kappa \varepsilon n_0 a^3\Delta a
\end{equation}
where $\varepsilon$ is the collision efficiency.
Setting $\Delta a=2.5\,\mu{\rm m}$ and $\varepsilon=0.03$ in addition to the parameters defined above
gives ${\cal R}\approx 10^{-4}{\rm s}^{-1}$. The rate of coalescence
of typical sized water droplets due to collisions is therefore very small.
Given that multiple collision events
are required to grow a droplet to the size where runaway growth is possible, explaining
growth by collisions is challenging.

Saffman and Turner \cite{Saf+56} investigated the role of turbulence in facilitating collisions
between water droplets. In the case of very small droplets, the collision rate due
to turbulence is a consequence of shearing motion, so that the collision speed is
of order $a_0/\tau_{\rm K}$. They argue that the corresponding collision rate is
\begin{equation}
\label{eq: 5.3}
{\cal R}_{\rm turb}=\sqrt{\frac{8\pi}{15}}\frac{n_0\varepsilon (2a)^3}{\tau_{\rm K}}
\ .
\end{equation}
For the parameters of the cloud model, the gives
${\cal R}_{\rm turb}\approx 2\times 10^{-6}\,{\rm s}^{-1}$,
which is negligible.

After a droplet has grown to a size where it is much larger than the typical droplets,
and where the collision efficiency is approximately unity, it falls rapidly and collects other droplets
in its path. Consider a droplet of size $a_1$ falling through a \lq gas' of small droplets,
which can be characterised by the liquid volume fraction $\Phi_{\rm l}=4\pi n_0\langle a^3\rangle/3$.
The large droplet falls with velocity $v=\kappa a_1^2$ and grows in volume at a rate $\pi a_1^2 \Phi_{\rm l}v$, so that
\begin{equation}
\label{eq: 5.4}
\frac{{\rm d}a_1}{{\rm d}t}=\frac{\kappa \Phi_{\rm l}a_1^2}{4}
\ .
\end{equation}
Solving this equation shows that the droplet radius diverges in the time
\begin{equation}
\label{eq: 5.5}
\tau_{\rm exp}=\frac{4}{\kappa \Phi_{\rm l}a_1}
\ .
\end{equation}
For the model parameters, a droplet of size $a_1=50\,\mu {\rm m}$
requires time $\tau_{\rm exp}\approx 2 \times 10^3\,{\rm s}$ to undergo explosive growth.
According to (\ref{eq: 5.5}) the time before runaway growth is expected to occur increases
rapidly as the droplet size gets smaller, and this estimate must be a lower bound because
it ignores the effects of collision efficiency and the settling velocity of the smaller droplets.

\section{Rare events and rain initiation}
\label{sec: 6}

Consider the rate at which droplets must reach the size threshold for runaway
growth. Rainfall at a rate of $3.6\,{\rm mm}\,{\rm hr}^{-1}=10^{-6}\,{\rm m}\,{\rm s}^{-1}$
is considered as \lq moderate'. If the raindrops have size $a\approx 1\,{\rm mm}$, this
corresponds to raindrops falling at a rate of approximately $250\,{\rm m}^{-2}{\rm s}^{-1}$.
Given the assumed cloud depth of $h=10^3\,{\rm m}$, the volumetric
rate of production of raindrops is approximately $0.25\, {\rm m}^{-3}{\rm s}^{-1}$. If the
microscopic droplets have density $n_0=4\times 10^8\,{\rm m}^{-3}$, then the rate of conversion
of each microscopic droplet into a \lq collector' droplet undergoing runaway growth is
approximately $6\times 10^{-10}\,{\rm s}^{-1}$.
An alternative statement is that if a shower lasts for a five minutes, the probability
that any given water droplet has grown to become a rain droplet is small,
approximately $2\times 10^{-7}$.
The problem of rain initiation is, therefore,
concerned with the frequency of very rare events. This point has also been made by Kostinski
and Shaw \cite{Kos+05}.

Growth of droplets from the typical size of $10\,\mu {\rm m}$ to $50\,\mu {\rm m}$
(which is the threshold for runaway) could in principle occur by collision and coalescence. However, despite the
fact that the required conversion probability is very small (of order $10^{-7}$), 
this is not achievable by a collisional mechanism.
On growing from $10\,\mu {\rm m}$
to $50\,\mu{\rm m}$, the volume of a droplet increases by a factor of $125$,
that is, there are or order $100$ collision events. The conversion of a droplet to become
a collector droplet requires a sequence of successive collisions which may be assumed
to be statistically independent. If the rates for successive collisions were all
equal to ${\cal R}$, the probability for $N$ collisions occurring after a short time $t$
would be
\begin{equation}
\label{eq: 6.1}
P_N\sim \frac{({\cal R} t)^N}{N!}
\end{equation}
It was argued above that the rate for the first collision events is small,
${\cal R}_0\approx 10^{-4}\,{\rm s}^{-1}$. Even allowing for the
fact that the collision rates increase as the droplet grows, the probability
for the obtaining $100$ collisions after $t=10^3\,{\rm s}$ will be much smaller
than $10^{-7}$. The collisional mechanism for bridging the bottleneck
to runaway growth is, therefore, highly problematic.

\section{Fast droplet growth by the convective mechanism}
\label{sec: 7}

As well as the theoretical difficulties of explaining droplet growth by collisional
processes, observational evidence is difficult to reconcile with a collisional
mechanism. Clouds may exist for long periods, before quite suddenly producing
rainfall. The rapid onset of rainfall is usually associated with convective
instability, which (because of the large Reynolds number) implies turbulent
motion. Equation (\ref{eq: 5.3}) indicates that the role of turbulence in
facilitating particle collisions is negligible for the small droplets in the
model treated here, implying that the rainfall is triggered by some other
aspect of the convective process. If a parcel of
air is lifted by convection, condensation occurs as the temperature falls.
However, the fractional increase of the droplet size which can be achieved is
not large enough to start runaway growth, and increasing the size of droplets by
condensation also reduces the dispersion of their radius.

For these reasons it is desirable to find other mechanisms whereby
convection can result in runaway growth. The Rayleigh-B\'enard cell can serve
as a model for convective motion in a cloud, and it will be argued that
the convective ripening mechanism can result droplet growth rates which
are more rapid than the collisional mechanism.

Droplets can grow or shrink due to changes in the level of supersaturation.
Consider the circulation of a droplet in a rising packet of air. This air mass is
cooled \emph{slowly} (on a timescale $\tau_{\rm c}\gg \tau_{\rm eq}$) by
radiation when it reaches the top of the cloud. The droplets that it
contains increase in size by condensation, due to capacity of the surrounding air
to carry water vapour being reduced. The cool packet of air then becomes part of a
\lq plume' of descending air, which falls far into the body of the cloud before
being \emph{rapidly} mixed with ambient air
(on a timescale $\tau_{\rm K}\ll \tau_{\rm eq}$). The temperature of the descending plume increases
due to adiabatic compression as the air pressure rises, but it is still colder than
the surrounding gas at the time when mixing occurs. The mixing occurs so rapidly that
the droplets are unable to evaporate, and their increased sizes are \lq frozen'.
If the droplets are close to the condensation level (the base of the cloud)
at the time when mixing
occurs, the droplets will achieve a size which is much larger than the
surrounding droplets. One single cycle of this process is not sufficient
to bridge the bottleneck and achieve runaway growth.
It will be argued that, compared to collisional processes, this mechanism can
require fewer steps for droplets to grow to the size where runaway growth is
possible, and that these steps can occur in a shorter timescale.
This mechanism requires $\tau_{\rm c}\gg \tau_{\rm eq}\gg \tau_{\rm K}$.
These inequalities are easily satisfied for the model cloud parameters, where
$\tau_{\rm c}\approx \tau_h\approx 10^3\,{\rm s}$, $\tau_{\rm eq}\approx 1\,{\rm s}$ and
$\tau_{\rm K}\approx 10^{-2}\,{\rm s}$.

Because $\tau_{\rm c}^{-1}$ is larger than the rate of collision of small particles,
the steps of the convective ripening mechanism are more frequent than those of the
collisional process. Recall that raindrops grow from a
very small number of microscopic droplets: in this case the shower
is triggered by droplets which happen
to be convected to the upper surface of the cloud several times in quick succession.

Equation (\ref{eq: 3.9}) implies that growth of water droplets by
condensation increases the area of a droplet by an amount which is
independent of the droplet size on each cycle. Consider what happens
as a droplet makes repeated encounters with the top of a cloud,
followed by rapid warming in the interior. Under the most favourable
circumstances, where a droplet falls repeatedly falls very close to the base of the cloud before its surroundings are mixed with the ambient air, the number of cycles required to increase the radius
from $10\,\mu {\rm m}$ to $50\,\mu{\rm m}$ is $(50/10)^2=25$. This is much smaller than
the number of events required for the collisional growth, which is $(50/10)^3\approx 125$, and the timescale
separating each event is shorter.

\section{Concluding remarks}
\label{sec: 8}

This paper has described a non-collisional model for increasing the dispersion
of droplet sizes in a Rayleigh-B\'enard cell. This is used as a model for
resolving the droplet growth bottleneck problem in cloud physics.
It has been argued above that the dominant mechanism for creating larger
droplets is that droplets grow slowly by condensation as they are convected
upwards in a cloud, but that the increased size is frozen in when a falling plume
of cold air is mixed rapidly in the interior of the cloud.

The convective ripening mechanism was compared with collisional growth
in clouds. The convective ripening mechanism discussed here can bridge the
growth bottleneck in fewer steps, which occur at a higher rate. Further
work in required to quantify the statistics of repeated contacts with
the cool plate of the convection cell, because this determines the rate
at which the largest droplets can grow.

\emph{Acknowledgements}. This paper was written with the generous support of the
NORDITA program \lq Dynamics of Particles in Flows'.

\end{document}